\documentclass[aps,prd,notitlepage,nofootinbib,superscriptaddress,
groupedaddress,showpacs]{revtex4}

\usepackage{graphicx,color}
\usepackage{amstext,amsmath,amssymb,amsfonts,bbm}
\usepackage{amsthm}
\usepackage{dcolumn}
\usepackage{bm}
\usepackage{hyperref}

\newcommand{\Ref}[1]{(\ref{#1})}

\newcommand{\Z}{\mathbb{Z}}

\newcommand{\C}{\mathbb{C}}

\newcommand{\half}{\frac{1}{2}}

\newcommand{\Tr}{\text{Tr}}

\newcommand{\ket}[1]{| #1 >}

\newcommand{\Slc}{\mathrm{SL}(2,\mathbb{C})}

\newcommand{\Su}{\mathrm{SU}(2)}

\def\be{\begin{eqnarray}}
\def\ee{\end{eqnarray}}

\renewcommand{\a}{\alpha}

\newcommand{\g}{\gamma}

\renewcommand{\L }{\Lambda}

\newcommand{\tr}{\mathrm{tr}}

\newcommand{\GI}[1]{(G_{{#1}}^{-1})^{\dagger}}

\def\bq{\begin{equation*}}
\def\eq{\end{equation*}}
\def\beq{\begin{equation}}
\def\eeq{\end{equation}}
\newcommand{\scal}[1]{\langle #1\rangle}
\usepackage[normalem]{ulem}
\renewcommand\sout{\bgroup\markoverwith{\textcolor{red}{\rule[0.5ex]{4pt}{0.8pt}}}\ULon}

\begin{document}

\title{Spin foam propagator: A new perspective to include the cosmological constant}

\author{Muxin Han}
\affiliation{Department of Physics, Florida Atlantic University, 777 Glades Road, Boca Raton, FL 33431, USA}
\affiliation{Institut f\"ur Quantengravitation, Universit\"at Erlangen-N\"urnberg, Staudtstr. 7/B2, 91058 Erlangen, Germany}

\author{Zichang Huang}\email{zhuang2014@fau.edu}
\affiliation{Department of Physics, Florida Atlantic University, 777 Glades Road, Boca Raton, FL 33431, USA}

\author{Antonia Zipfel}
\affiliation{Department of Physics, Florida Atlantic University, 777 Glades Road, Boca Raton, FL 33431, USA}

\date{\today}
\begin{abstract}
In recent years, the calculation of the first non-vanishing order of the metric 2-point function or graviton propagator in a semiclassical limit has evolved as a standard test for the credibility of a proposed spin foam model. The existing results of spinfoam graviton propagator rely heavily on the so-called double scaling limit where spins $j$ are large and the Barbero-Immirzi parameter $\gamma$ is small such that the area $A\propto j\gamma$ is approximately constant. However, it seems that this double scaling limit is bound to break down in models including a cosmological constant. We explore this in detail for the recently proposed model \cite{Haggard:2014xoa} by Haggard, Han, Kaminski and Riello and discuss alternative definitions of a graviton propagator, in which the double scaling limit can be avoided.

\end{abstract}

\maketitle

\section{Introduction}

Spin foam models \cite{carlobook,Perez:2012wv} aim at a path integral description of Loop Quantum Gravity (LQG). Despite tremendous developments in recent years, most models struggle in including a cosmological constant $\Lambda$. Yet, the empirical data clearly hints at a non-vanishing, positive cosmological constant.  In order to develop more realistic models, it is therefore of great importance to incorporate a cosmological constant. As shown in e.g. \cite{Smolin:2002sz,Han:2011aa,Ding:2011hp, Bianchi:2011uq} a cosmological constant might also serve as a natural regulator through a quantum group structure, which has been established in Euclidean spin foam models. For Lorentzian signature the connection between a quantum group and a cosmological constant is so far unknown. But an alternative approach towards including a cosmological constant  has been suggested in \cite{Haggard:2014xoa,Haggard:2015ima, Haggard:2015kew,Haggard:2015yda}. The guiding idea of \cite{Haggard:2014xoa,Haggard:2015ima, Haggard:2015kew,Haggard:2015yda} is to express the Lorentzian spin foam action with cosmological constant as a 
$\Slc$-Chern-Simons theory evaluated on a specific graph observable $\Gamma_5$ (see Figure \ref{fig:g5}), which can be interpreted as the dual graph of a constantly curved 4-simplex.

Due to the lack of experimental and observational data, testing the semiclassical properties of a proposed model of quantum gravity is crucial to justify the assumptions made. In spin foam models there exist essentially two standard test of this kind: On the one hand, the spin foam amplitude of a semiclassical state is governed by a phase depending on the discrete Regge action in the limit where spins are large (see e.g. \cite{Barrett:2009mw,Barrett:2009gg,HZ,HZ1,CFsemiclassical}). As shown in \cite{Haggard:2014xoa}, the model  \cite{Haggard:2014xoa,Haggard:2015ima, Haggard:2015kew,Haggard:2015yda} reproduces the correct Regge-phase with cosmological constant in a semiclassical limit where spins $j$ (i.e. areas) and the Chern-Simons coupling $|h|$ become large w.r.t. $\hbar$. On the other hand, the first non-vanishing order of the spinfoam graviton n-point function should reproduce the one of Regge calculus in the semi-classical limit.
 In fact, it was this latter test (see \cite{Alesci:2007tx, Alesci:2007tg}) that revealed the shortcomings of the Barrett-Crane model \cite{Barrett:1997gw, Barrett:1999qw} and led to the development of the Engle-Pereira-Rovelli-Levine (EPRL) model  \cite{Engle:2007qf, Engle:2007wy,Pereira:2007nh}. \footnote{See e.g. \cite{Alesci:2007tx, Alesci:2007tg, Bianchi:2009ri, Bianchi:2011hp, Shirazi:2015hwp,Rovelli:2011kf} for the recent results on spinfoam graviton propagator and 3-point function.} The aim of this paper is to establish the graviton propagator for the model with cosmological constant\cite{Haggard:2014xoa,Haggard:2015ima,Haggard:2015kew,Haggard:2015yda}.

Although semi-classical behavior of spinfoams is expected to be achieved in the large spin limit, the existing results on graviton n-point function requires more input in taking the limit. Namely, it requires a double scaling limit to recover the semi-classical graviton n-point function. The double scaling limit consist of taking large spins $j$ but at the same time small Barbero-Immirzi parameter $\gamma$, such that the kinematical area $A\approx \gamma j$ stays constant. While this limit works fine in models without cosmological constant (see e.g. \cite{Alesci:2007tx, Alesci:2007tg,Bianchi:2009ri, Bianchi:2011hp, Shirazi:2015hwp,Han:2013ina,Magliaro:2011zz,Magliaro:2011dz}) it is bound to break in models that include a cosmological constant. 
This can be easily seen from the following argument: In the spinfoam model with cosmological constant\cite{Haggard:2014xoa,Haggard:2015ima, Haggard:2015kew,Haggard:2015yda,4dqg,Han:2017geu}, one has to take an additional limit such that $\L\to0$ in the same rate as $j$ becomes large in order to recover the correct semiclassical behavior, i.e. the Regge action on a constantly curved 4-simplex. Then if one takes additionally $\g\to0$, in the 4-simplex Regge action $\sum_f \g j_f\Theta_f+\L V_4$ \footnote{$f$ denotes a triangle in the 4-simplex. $\Theta_f$ denotes a dihedral angle.}, the two terms scale differently. At least when the cosmological constant is small, the 4-simplex volume $V_4$ behaves as $\gamma^{2}j^2$ while the area is $\gamma j$, so both don't scale ($\Theta_f$ doesn't scale). But $\L$ scales to zero, which eliminates the cosmological constant term in the first non-vanishing order of the graviton propagator. 

However, in the standard proposal of spinfoam propagator which employs a particular choice of the metric operator, the double-scaling limit is necessary in order to suppress non-classical contributions in the first non-vanishing order  (see e.g. \cite{Alesci:2007tx, Alesci:2007tg,Bianchi:2009ri, Bianchi:2011hp, Shirazi:2015hwp} ). As the calculations in section \ref{sec:lim} reveals, this is also the case in the model \cite{Haggard:2014xoa} if one follows the standard proposal. Yet, as argued above, sending $\gamma$ to zero will eliminate also the cosmological constant term. To resolve this dilemma, it seems necessary to reconsider the definition of the graviton propagator. 

In the traditional approach  \cite{Alesci:2007tx, Alesci:2007tg,Bianchi:2009ri, Bianchi:2011hp, Shirazi:2015hwp} the propagator is constructed out of the metric operator of canonical Loop Quantum Gravity \cite{book,review1,review,book1}. While this is certainly a viable choice it posses several questions. Firstly, there exists no proof that canonical and covariant LQG are compatible in the sense that operators of the canonical theory can be directly mapped to operators in spin foam models. 
On the other hand, the metric in the canonical theory is defined on the kinematical level. But spin foam models supposedly solve all the constraint and therefore should yield the expectation values of physical not kinematical operators. For these reasons one might very well consider a metric operator that is directly adapted to the spin foam setting and does not make immediate use of the canonical theory. This point of view is in particular supported if spin foam models are interpreted as truncated theory theories for discrete quantum gravity, whose relation to a full theory of quantum gravity can only be recovered in a continuum limit. Understanding spinfoam graviton propagator in the semiclassical continuum limit is a research undergoing, based on the recent result in \cite{Han:2017xwo}.

We suggest to replace the metric in the propagator by an operator that only depends on the spins. The operator is only defined locally in the parameter space of the boundary data. In other words, the operator is specifically tied to the boundary state and a neighborhood in the parameter space of the boundary data. It is natural from the perturbative QFT perspective in which the perturbative QFT operators are usually defined upon a choice of vacuum of the theory. Here in the spinfoam amplitude, the boundary state plays the role of a vacuum state for a perturbation theory over the geometry defined by the boundary state. By a simple argument it can be shown that the limit $\gamma\to 0$ becomes superfluous for so-constructed operators. This solves the problems discussed above and enables to implement a cosmological constant.

In section \ref{sec:propagator} we will review the original construction of the graviton propagator and discuss alternatives directly adapted to spin foam models. These different choices for a graviton propagator are then analyzed in the context of the recently proposed model with cosmological constant \cite{Haggard:2014xoa}, which we will briefly review in section \ref{sec:mod}. As shown in \cite{Haggard:2014xoa}, the model reproduces the correct Regge-phase with cosmological constant in a semiclassical limit where spins $j$ (i.e. areas) and the Chern-Simons coupling $|h|$ become large w.r.t. $\hbar$. It is, hence, ideally suited to demonstrate the problems of the double scaling limit in the presence of a cosmological constant. As shown in section \ref{sec:lim}, the expected semiclassical result can only be reproduced for the modified graviton propagator and not for the original one, giving further evidence that a different construction of the propagator might be necessary. The paper concludes with a discussion of these findings in section  \ref{sec:dis}.

\section{Different proposals for a graviton propagator}
\label{sec:propagator}
\subsection{Standard proposal and conflicts with a cosmological constant }
Recall that a spin foam amplitude provides a map from the  Hilbert space ${\cal H}_{\partial {\cal R}}$ induced on the discrete boundary $\partial {\cal R}$ of a region ${\cal R}$ into $\C$. That is,
$W: \Phi \mapsto \scal{W|\Phi}$  for $ \Phi\in{\cal H}_{\partial {\cal R}}$. 
The expectation value of an observable ${\cal O}$, in the sense of the general boundary proposal \cite{Conrady:2003sw},  is then given by
\begin{equation}
\label{eq:exp1}
\scal{\mathcal{O}}=\frac{\scal{W|\mathcal{O}|\Phi}}{\scal{W|\Phi}}~.
\end{equation} 
Thus the metric 2-point function or graviton propagator is of the form 
\begin{equation}
\label{eq:2f}
G^{\alpha\beta\gamma\delta}(x,y)=\scal{q^{\alpha\beta}(x)q^{\gamma\delta}(y)} - \scal{q^{\alpha\beta}(x)}\scal{q^{\gamma\delta}(y)}~.
\end{equation}
Note that we are here working with  rescaled inverse density-two metric  $q^{\alpha\beta}=\det h\; h^{\alpha\beta}$ rather than the boundary metric $h_{\alpha \beta}$ on $\partial{\cal R}$ in order to allow for a direct comparison with canonical LQG. As any operator in LQG,  $q^{\alpha\beta}$ must be regularized. Since in the first order formalism $q^{\alpha\beta}$ is obtained by contracting the densitized co-triads $E^{\alpha}_i $, i.e. $q^{\alpha\beta}= E^{\alpha}_i E^{\beta i}$, $q^{\alpha\beta}$ will be smeared over the surfaces dual to the edges of the graph $\gamma$ over which the boundary state $\Phi$ is defined. In the following, we will restrict to graphs $\Gamma_5$ that are dual to a 4-simplex $\sigma$, since the 4-simplex amplitude is the most fundamental one in all spin foam models. In this setup, the discretized metric at the node $n$ is of the form 
\bq
q_n^{ab} := E^{a}_i(n) E^{b}_i(n)
\eq
where $E^{a}_i(n)$ is the co-triad smeared over the triangle $\Delta_{na}$ in $\sigma$ that is shared by the tetrahedra $\tau_n$ and $\tau_a$. It follows that the discrete graviton propagator can be generically written as 
\beq
\label{eq:2pt1}
G^{abcd}_{mn}=\scal{q^{ab}_n \ q^{cd}_m }-\scal{q^{ab}_n}\scal{q^{cd}_m } ~.
\eeq  
In older approaches the co-triads $E^{a}_i(n)$  are replaced by the flux operators of canonical LQG, i.e. they act as the right invariant vector fields on the edges of the boundary spin network (see e.g  \cite{Rovelli:2005yj,Bianchi:2006uf,Alesci:2007tx,Bianchi:2009ri} for details). The first non-vanishing order in these approaches is then found by performing an asymptotic analysis for large spins, that is  $j \to \lambda j$ for $\lambda>>1$, and takes the generic form
\beq
\label{eq:gen_prop}
G(\lambda)\approx (\gamma\lambda)^3 q_{,j} \,q_{,j'}\,(H_{Regge})_{jj'}^{-1} +\gamma^4\lambda^3(\hat{H}+\mathcal{O}(\gamma)),
\eeq 
where $H_{Regge}$ is the Hessian of the Regge action\footnote{without cosmological constant and $\gamma=1$} as a function of $j$, where $q_{,j}$ is the derivative of the expectation value of $q^{ab}_n$ with respect to $j$  and where  $\hat{H}$ is independent of $\gamma$. In order to suppress the non-classical term proportional to $\hat{H}$ previous works now enforce the additional limit $\gamma \to 0$ keeping the area $A={\gamma}{l_p^2}\sqrt{j(j+1)}$ approximately constant.

For models with a cosmological constant we  expect a similar result with the difference that the Hessian now depends on the  action 
\begin{equation}
\label{eq:rega}
S_{Regge}=-\frac{i}{l_p^2}\left[\sum_{(ab)}A_{ab}\Theta_{ab}-\Lambda V_4\right],
\end{equation} 
where $A_{ab}$ is the area, $\Theta_{ab}$ stands for the dihedral angle, and where $V_4$ is the 4d volume. Indeed, this is exactly what we did find for the recently proposed  model by Haggard, Han, Riello and Kaminski \cite{Haggard:2014xoa,Haggard:2015ima, Haggard:2015kew,Haggard:2015yda} if one considers instead of the pure large $j$-limit the double scaling limit  $j\to\lambda j$ and $\Lambda\to\frac{\Lambda}{\lambda}$ (see section \ref{sec:lim} for details). But now the limit $\gamma\to 0$ can no longer be considered since the Regge action is no longer linear in $\gamma$. While the area is linear in $\gamma$ the 4d volume scales as area squared and, hence, depends quadratically on $\gamma$. Consequently, the area term would be much greater than the volume term in the limit $\gamma\to 0$, which would suppress the cosmological constant term in the Regge action as well as in the Hessian. This is obviously not what we expect. 

The above considerations suggest that there is a generic problem in deriving the graviton propagator when a cosmological constant is included, which is not restricted to the model \cite{Haggard:2014xoa,Haggard:2015ima, Haggard:2015kew,Haggard:2015yda} analyzed in greater detail in the subsequent section. Consequently, we should revisit the construction of the graviton propagator itself. Recall that the densitized co-triad $E^\a_i$ is defined on the kinematic level since it originates from the 3+1 decomposition before the Hamiltonian constraint is applied. But the 2-point function should yield the expectation value for an incoming and an outgoing graviton excitation on a coherent boundary state on the dynamic level. Moreover, there is no formal proof that canonical and covariant LQG are compatible in the sense that operators carry-over from canonical to covariant LQG. So, it is not a priori clear whether the canonical flux operators are the only viable choice to define the metric 2-point function. Instead one could choose an approach in which the metric operator is based on variables that are more inherently defined in the spin foam model. 
\subsection{Perturbative truncated metric}
The most promising candidate, which can solve the problems mentioned above, is a metric operator $q(j)$ that only depends on the area-variables, i.e. spins $j$. Since 
the only non-zero derivative with respect the system variables is in this case $q(j)_{,j}$,  the first non-vanishing order of the asymptotic expansion \eqref{eq:gen_prop} 
takes the form
\begin{equation}\label{eq:gj}
G(\lambda)\approx\lambda^3 q_{,j} \,q_{,j'}\,(H_{Regge})_{jj'}^{-1},
\end{equation}
which only contains the expected term. 
%
Since areas and surface normal determine a 4-simplex uniquely up to translation and inversion, an example of a metric operator in the above scenario can be 
\begin{equation}
\label{eqn:ex_metric}
(q_{\xi})^n_{ab}=\delta_{ij}(\gamma j_{na}n^i_{na})(\gamma j_{nb}n^j_{nb})
\end{equation}
where $n^i_{ab}$ is the normal to the triangle $\Delta_{ab}$. 
	
This choice might be too simple 
in the sense that it depends heavily on the normals, which are fixed by the boundary data. 
A less trivial proposal is to express the edge-lengths in terms of the area variables and construct the metric by those edge-area relations.

Since a 4-simplex is uniquely fixed by 10 independent edge lengths, the discrete metric is also uniquely determined by those lengths.  In particular, this means that we can determine the normals $n_{ab}^i$ in \eqref{eqn:ex_metric}, which are related to the discretized co-triads, as functions of the lengths. On the other hand, there are exactly 10 areas in a 4-simplex, so that it looks tempting to express the metric in terms of the areas by solving the inverse of the Heron formula \eqref{eq:heron} and the corresponding 4-simplex constraints.
Heron's formula \footnote{Heron's formula works for the flat tetrahedron. In the spherical case, the area can be determined by the edge-lengths through L'Huilier's theorem. In hyperbolic case, one can also get a similar relation through the hyperbolic law of cosine. The details are discussed in the appendix.\ref{app:area}. } is given by
\begin{equation}\label{eq:heron}
A_{ij}=\frac{1}{4}\sqrt{(l_k+l_l+l_m)(-l_k+l_l+l_m)(l_k-l_l+l_m)(l_k+l_l-l_m)},
\end{equation}
where $A_{ij}$ stands for the area constructed by the edges $l_k,l_l,l_m$.
%
%
But, as a second order equation, the inverse of Heron's formula has more than one solution and the solution of the full system of equations is therefore ambiguous (see e.g. \cite{Barrett:1997tx} ). 
In fact, expressing the metric purely in terms of areas faces the same problem as earlier attempts to define the Regge action in terms of areas (see e.g. \cite{Rovelli:1993kc,Makela:1998wi}), namely that the areas are 
subjected to hidden constraints (see e.g. \cite{ Barrett:1997tx}\cite{Makela:2000ej}\cite{ Regge:2000wu}\cite{Wainwright:2004yn}). Thus, we also need to consider variables fixed by the boundary state, e.g. the dihedral angles\footnote{This is also a viable choice for curved simplices, see \cite{Bahr:2009qd}} as suggested in \cite{Dittrich:2008va}. This is possible since the ambiguity mentioned in \cite{Barrett:1997tx} is discrete.  For ten given areas there are multiple choices of the edge lengths to reconstruct a 4-simplex but there is no continues deformation between these choices. For a chosen set of edge-lengths, the perturbation on the geometry cannot transform itself into another 4-simplex geometry that match the same areas. More specifically, for a fixed boundary state, the solution of the inverse of the Heron's formula must be uniquely chosen to match the edge-lengths of the boundary state. Under this circumstance, we can construct the metric as a function of the area variables which are valued in the neighborhood of the exact areas given by the edges lengths of the fixed boundary state. Within the neighborhood of the given areas, the variation of the areas will not change the choice of the inverse solution of the Heron's formula. So the exact form of the metric function will also {remain}. However, a so-constructed metric is only locally defined in the parameter space of boundary data since it is only valid for a specific choice of boundary data.

In order to show 
that it is still sensible to define a propagator by using a boundary-data-dependent metric operator as described above,  
let us briefly revisit the generalized boundary proposal \cite{Conrady:2003sw} underlying the construction of the graviton propagator. As pointed out in \cite{Bianchi:2006uf}, there is no preferred vacuum state in a background-independent quantum gravity theory. Instead, the 2-point function is evaluated on a specific geometry encoded in the boundary states, which can be interpreted as the `vacuum state' around which we are considering quantum perturbations. This means that for each different boundary geometry, we obtain a different 2-point function as a result of a perturbative truncated field theory. 
For such a perturbatively defined truncated field theory, we may therefore consider a metric operator in the above sense. The metric operator is defined upon a choice of the vacuum state on which the perturbation theory is defined.
%
 More specifically, within a truncated field theory, defined by a specific boundary state $\ket{\Phi}_{j}$, the metric operator $\hat{q}_{j}$ defined by the area variables $j$ lead to the following 2-point function: 
\begin{equation}
\label{eq:2pt1j}
(G_j)^{abcd}_{nm}=\scal{(q_j)^{ab}_n \,(q_j)^{cd}_m} - \scal{(q_j)^{ab}_n}\scal{(q_j)^{cd}_m}_j~.
\end{equation}
Here $(q_j)^{ab}_n$ is given by Eq.\Ref{eqn:ex_metric}, while viewing $n^i_{na},n^j_{nb}$ as functions of $j$. Then the truncated expectation value is 
\begin{equation}
\label{eq:exp1j}
\scal{\mathcal{O}}_j=\frac{\scal{W|\mathcal{O}|\Phi}_j}{\scal{W|\Phi}_j}~.
\end{equation} 
From \eqref{eq:gj} it then follows immediately that the first non-vanishing order of the asymptotic expansion of \eqref{eq:2pt1j} matches the expected Regge-like form.



As shown in section \ref{sec:lim}, the second term in \eqref{eq:gen_prop} vanishes if the metric operator only depends on the areas since $\hat{H}$ depends on the derivatives of $q$ with respect to the variables distinct from $j$. Consequently, the limit $\gamma\to0$ becomes superfluous. The same argumentation is also valid for constantly curved simplices in the model \cite{Haggard:2014xoa,Haggard:2015ima, Haggard:2015kew,Haggard:2015yda}. Note that the boundary data in this model fixes the sign of the cosmological constant, which determines whether the 4-simplex is spherical or hyperbolic.   Furthermore, each face of the 4-simplex is flatly embedded in an ambient space $\mathcal{S}^3$ or $\mathcal{H}^3$ \cite{Haggard:2015ima}, so that it is still possible to construct an unique metric operator from the area variables for a fixed boundary. The asymptotic analysis is then very similar to the flat case and will be discussed in the rest of this article.


\section{Spin foam model with cosmological constant}
\label{sec:mod}

\begin{figure}
	\centering\includegraphics{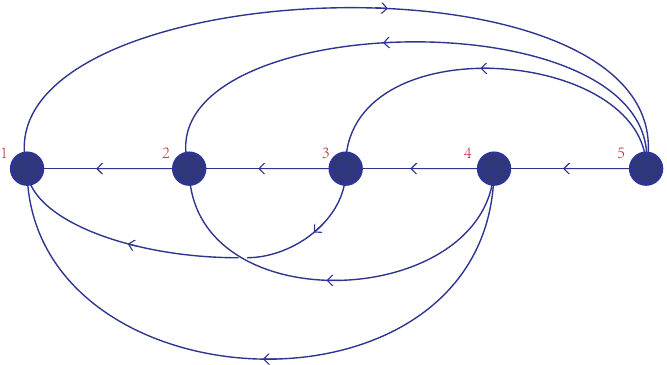}
	\caption{The dual graph $\Gamma_5$ of a 4-simplex lives on the (spatial) boundary, i.e. in $S^3$. It's vertices are dual to the five tetrahedra of the 4-simplex and its edges are dual to the triangles. The figure above depicts a projection of $\Gamma_5$ into the plane.}
		\label{fig:g5}

\end{figure}

Before turning to the analysis of the various graviton propagators let us briefly review the model \cite{Haggard:2014xoa,Haggard:2015ima, Haggard:2015kew,Haggard:2015yda} with which we are working. The model can be derived from the action \footnote{$\prec \cdot ,\cdot \succ$ is the invariant, non-degenerate bilinear form of $\mathfrak{sl}(2,\mathbb{C})$ which couples boost and rotation. For example $\prec X,Y \succ$ is $\frac{1}{2}\epsilon_{IJ}^{\ \ KL}X^{IJ}Y_{KL}$.}(see \cite{Haggard:2014xoa} for details)
\beq
\label{eqn:lambdaBF}
{\cal S}_{\Lambda BF}:= -\frac{1}{2} \int_{\cal M} \prec\left[\left(1-\frac{1}{\gamma} \star\right) B \right] \wedge F[A] - \frac{\Lambda}{6} \left[\left(1-\frac{1}{\gamma} \star\right) B \right] \wedge B \succ~,
\eeq
where $B$ is a bivector, $F[A]$ is the curvature of an $\Slc$-connection $A$ on space-time ${\cal M}$ and $\gamma$ is the Barbero-Immirzi parameter. This reduces to the gravitational Holst-action once the so-called simplicity constraint $B=e\wedge e$ is imposed, with $e$ being a tetrad on ${\cal M}$. Following the general strategy in modern spin foam models \cite{Engle:2007qf, Engle:2007wy,Pereira:2007nh}, one now derives the path integral of \eqref{eqn:lambdaBF} on a given (simplicial) discretization of ${\cal M}$. In the following, we restrict our attention to a single curved 4-simplex\footnote{In contrast to other spin foam models we here consider constantly curved simplices instead of piecewise linear ones. Edges and faces of the tetrahedra are flatly embedded into $S^3$ or $\mathcal{H}^3$ depending on the sign of $\Lambda$ (see \cite{Haggard:2015ima} for details).} and its dual graph $\Gamma_5$. The amplitude associated to a boundary state $\Psi_{\Gamma_5}$ is given by 
\bq
\scal{\Lambda BF |\psi_{\Gamma_5}} =
\int {\frak D}A\,{\frak D}\Pi \exp(-i{\cal S}_{\Lambda BF})\, \psi_{\Gamma_5}~,
\eq  
where ${\frak D}$ denotes the path integral measure and $\Pi=\left(1-\frac{1}{\gamma} \star\right) B$. Integrating out $\Pi$ and splitting into self- and anti-self-dual part yields
\beq
\label{eqn:LBF_amp}
\scal{\Lambda BF |\psi_{\Gamma_5}}= 
\int {\frak D}A\,{\frak D}\bar{A}\, \exp\left(-i \frac{h}{2} W[A] -i\frac{\bar{h}}{2} W[\bar{A}] \right)\psi_{\Gamma_5}[A,\bar{A}]
\eeq
with
\bq
W[A]:=\frac{1}{4\pi} \int_{S^3} \tr(A\wedge dA+\frac{2}{3}A\wedge A\wedge A)
\eq
and $h=\frac{12 \pi}{\Lambda}\left(\frac{1}{\gamma} +i\right)$. The simplicity constraint is now imposed on the boundary spin network, restricting $\psi_{\Gamma_5}$ to a state $\Gamma_5(\vec{j},\vec{i}| A,\bar{A})$  solely labeled by $\Su$-spin-network data $(\vec{j},\vec{i})$ in exactly the same manner as in the Engle, Pereira, Rovelli, Levine (EPRL) model \cite{Engle:2007qf, Engle:2007wy,Pereira:2007nh}. If we choose coherent boundary data, i.e. Levine -Speziale intertwiners $\vec{i}_{LS}$ then  $\Gamma_5(\vec{j},\vec{i}_{LS}| A,\bar{A})$ is given by 
\beq 
\label{eqn:graph-obs}
\begin{split}
\Gamma_5(\vec{j},\vec{i}_{LS}| A,\bar{A})
&=\int_{\Slc} \prod_{a=1}^5 d g_a \prod_{a<b} P_{ab}(g_a,g_b,G_{ab})
\\
&:=
\int_{\Slc} \prod_{a=1}^5 d g_a \prod_{a<b} \scal{j_{ab},-\vec{n}_{ab}|Y^{\dagger}g_a^{-1} \,G_{ab}[A,\bar A]\, g_b\,Y|j_{ab},\vec{n}_{ba}}~.
\end{split}
\eeq 
Here, $a$ label the vertices of $\Gamma_5$ or tetrahedra $\tau$ of $\sigma$ respectively, $\vec{n}_{ab}$ is the normal associated to the triangle $\Delta_{ab}$ as seen from $\tau_a$ \footnote{In the piecewise linear setting $\vec{n}_{ab}$ is the outward pointing normal of $\Delta_{ab}$. In the case of a constantly curved tetrahedron the closure condition is given by a cyclic conditions on the holonomies $h_{ab}$ starting at a base point and encircling  $\Delta_{ab}$. Since $\Delta_{ab}$ are flatly embedded surfaces these holonomies are completely determined by the normals $\vec{n}_{ab}$ at the base point and the areas of $\Delta_{ab}$. See \cite{Haggard:2015ima} for details.}
and $|j,\vec{n}\rangle$ represent the coherent states defined by Perelomov \cite{Perelomov:cs1972,PerelomovGCS}. Furthermore, $Y$ is the EPRL-map \cite{Pereira:2007nh}, that maps the $\Su$-irreducible ${\cal H}_j$ into the $\Slc$-irreducible ${\cal H}^{j,\gamma j}$, and $G[A,\bar{A}]$ denotes the holonomy of the Chern-Simons connection. 
Since the Chern-Simons term in \eqref{eqn:LBF_amp} is invariant under local $\Slc$ transformations (modulo $2\pi\Z$) and \eqref{eqn:graph-obs} is invariant under the gauge transformation
\begin{equation}
\label{eq:abc}
g_a \mapsto G(a)g_a\ ;G_{ab}\mapsto G(a)G_{ab}G(b)^{-1}
\end{equation}
we can fix $g_a=\mathbbm{1}$ and drop the infinite integral $\int\prod d g_a$ so that the $\Lambda$-deformed EPRL-amplitude \cite{Haggard:2014xoa} is finally given by
\beq
\label{eq:HAN1}
 \scal{W_{\Lambda EPRL}|\sigma; \vec{j},\vec{n}}=\int \mathfrak{D}A\mathfrak{D}\bar{A} \exp\left(-i \frac{h}{2} W[A] -i\frac{\bar{h}}{2} W[\bar{A}] \right)\prod_{a<b} P_{ab}(G_{ab})~.
\eeq,
where $P_{ab}(G_{ab})$ can be expressed as
\beq
P_{ab}(G_{ab})=\scal{j_{ab},-\vec{n}_{ab}|Y^{\dagger} \,G_{ab}[A,\bar A]\, \,Y|j_{ab},\vec{n}_{ba}}.
\eeq
Following \cite{Haggard:2014xoa,Barrett:2009mw}, $P_{ab}$ can be expressed as exponential by using the representation of the coherent states $Y|j,\vec{n}\rangle$ in terms of homogeneous functions $f(z)$ of two complex variables $(z_0,z_1):=z$ on which $g\in\Slc$ acts by its transpose (i.e. $g f(z)=f(g^T z)$). More precisely, 
\bq
\scal{z|Y|j,\xi} :=f_{\xi}^{j}(z)^{(j,\gamma j)}=\sqrt{\frac{(2j+1)}{\pi}}<z,z>^{i\gamma j-1-j}<\bar{z},\xi>^{2j},
\eq  
where $\xi$ is the unit spinor associated to $\vec{n}$ \footnote{ I.e. $\scal{\xi|\vec{\sigma}\xi}=\vec{n}(\xi)$ where $\sigma^i$ are the Pauli matrices},  and $J$  maps $(z_0,z_1)^T$ to $(-\bar{z}_1,\bar{z}_0)^T$. This implies
\begin{equation}\label{eq:HAN3}
\begin{split}
P_{(ab)}
=\int_{\mathbb{CP}^1}d\mathbf{z}_{ab}\;\overline{f^{j_{ab}}_{J\xi_{ab}}(z_{ab})^{(j_{ab},\gamma j_{ab})}}\;f^{j_{ab}}_{\xi_{ba}}(G_{ab}^{T}z_{ab})^{(j_{ab},\gamma j_{ab})}
\end{split}
\end{equation}
with measure $d\mathbf{z}=\frac{i}{2}(z_0dz_1-z_1dz_0)\wedge(\bar z_0d\bar z_1-\bar z_1d \bar z_0)$. After a further change of variables, $z\to G^{T} z$ and $z\to \bar{z}$, which is introduced for later convenience, we therefore obtain\footnote{See equation (5.20) in \cite{Haggard:2014xoa}}
\beq\label{eq:HANF}
 \scal{W_{\Lambda EPRL}|\sigma; \vec{j},\vec{n}}= 
\int\mathfrak{D}A\,\mathfrak{D}\bar A\int_{(\mathbb{CP}^1)^{10}}\!d\mu(z)\; e^{S[A,\bar{A},\vec{z};\vec{j},\vec{\xi}]}
\eeq
with
\begin{widetext}
\begin{equation}\label{eq:act1}
\begin{split}
S[A,\bar{A},\vec{z};\vec{j},\vec{\xi}]
=&-i\frac{h}{2}W[A]-i\frac{\bar h}{2} W[\bar A]+\sum_{(ab)} 2j_{ab} \ln{\frac{<J\xi_{ab},\GI{ab}z_{ab}><z_{ab},\xi_{ba}>}{<\GI{ab}z_{ab},\GI{ab}z_{ab}>^{\frac{1}{2}}<z_{ab},z_{ab}>^{\frac{1}{2}}}}\\
&+i\gamma j_{ab} \ln{\frac{<z_{ab},z_{ab}>}{<\GI{ab}z_{ab},\GI{ab}z_{ab}>}}
\end{split}
\end{equation}
\end{widetext}
and
\bq
d\mu(z)= \prod_{(ab)}\frac{-(2j_{ab}+1)}{\pi \|z_{ab}\|^2  \|\GI{ab}z_{ab}\|^2}d\mathbf{z}_{ab}~.
\eq

\subsection{2-point function}
\label{sec:2-p}

As discussed in section \ref{sec:propagator}, the metric operator in the traditional approach \cite{Rovelli:2005yj,Bianchi:2006uf,Alesci:2007tx,Bianchi:2009ri} is constructed from the canonical flux operator $E^a_b$ which act as the right invariant vector fields on the links $(ab)$. That is, for a single link in \eqref{eqn:graph-obs} one finds (analogously to \cite{Bianchi:2011hp}):
\begin{widetext}
\begin{equation}
\label{eq:2ptq}
\begin{split}
&\scal{j_{ab},-\vec{n}_{ab}(\xi)|Y^{\dagger}G_{ab}Y(E^a_b)^i|j_{ab},\vec{n}_{ba}}\\
&\qquad=\frac{2j_{ab}+1}{\pi}\int d\tilde{\mathbf{z}}_{ab}
\left(\frac{\|z_{ab}\|}{\|\GI{ab}z_{ab}\|}\right)^{2i\gamma j_{ab}}\left(\frac{\scal{J\xi_{ab},\GI{ab}z_{ab}}^2\scal{z_{ab},\xi_{ba}}^2}{\scal{\GI{ab}z_{ab},\GI{ab}z_{ab}}\scal{z_{ab},z_{ab}}}\right)^{j_{ab}}
\;A^i_{ab}(j_{ab},z_{ab})
\end{split}
\end{equation}
\end{widetext}
with $A^i_{ab}(j_{ab},z_{ab})=\gamma j_{ab}\frac{\scal{\sigma^iz_{ab},\xi_{ba}}}{\scal{z_{ab},\xi_{ba}}}$ and $\tau^i=\frac{\sigma^i}{2i}$.  Note that the same is true for the new metric operator \eqref{eqn:ex_metric} discussed in section \ref{sec:propagator} only that $A^i_{ab}$ now depends solely on $j_{ab}$ and not on $z_{ab}$. Thus in both cases we find:
\begin{gather}
\label{eq:2pt3}
\begin{gathered}
 \scal{W_{\Lambda EPRL}|E^a_n \cdot E^b_n|\sigma; \vec{j},\vec{n}}
=\int\mathfrak{D}A\,\mathfrak{D}\bar A \,d\mu(z)\,e^{S}\,
\delta_{ij}\,A^i_{na}\,A^j_{nb}\\
 \scal{W_{\Lambda EPRL}|E^a_n \cdot E^b_n\ E^c_m \cdot E^d_m|\sigma; \vec{j},\vec{n}}
 =\int\mathfrak{D}A\,\mathfrak{D}\bar A\,d\mu(z)\,e^{S}\,  \delta_{ij}\,A^i_{na}\,A^j_{nb}\,\delta_{kl}\,A^k_{mc}\,A^l_{md}~.
\end{gathered}
\end{gather}
The 2-point function \eqref{eq:2pt1} depends of course on the choice of the boundary state. In order to test the semiclassical properties of \eqref{eq:2pt1} it is therefore important to choose a state with appropriate semiclassical properties. As advertised in \cite{Rovelli:2005yj}, a good choice\footnote{These states are closely related to the so-called complexifier coherent states discussed in e.g.  \cite{Bahr:2007xa,Bahr:2007xn}. Also see \cite{Bianchi:2009ky}.}, that is peaked on intrinsic and extrinsic geometry alike, is given by a superposition of the states $|\sigma; \vec{j},\vec{n}\rangle$ of the form
\begin{widetext}
\begin{equation}
\label{eq:semis}
\begin{split}
|\Phi_0\rangle&=\sum_{\vec{j}}\Phi_{\vec{j},\vec{j_0}} |\sigma; \vec{j},\vec{n}\rangle\\
&=\sum_{\vec{j}} \exp\left[-\sum_{(ab)(cd)}\gamma\alpha_{(ab)(cd)}\frac{(j_{ab}-j_{0ab})}{\sqrt{j_{0ab}}}\frac{(j_{cd}-j_{0cd})}{\sqrt{j_{0cd}}}-i\sum_{(ab)}(\gamma\phi^0_{ab}(j_{ab}-j_{0ab}) - j_{ab}\theta_{ab})\right]
|\sigma; \vec{j},\vec{n}\rangle~.
\end{split}
\end{equation}
\end{widetext}
Here, $\phi^0_{ab}$ are the dihedral angles of the tetrahedra, which encode the extrinsic curvature\footnote{see section e.g. section 10.2 of \cite{Haggard:2014xoa} for the detailed construction}, and $\alpha_{(ab)(cd)}$ is a complex $10\times 10$ matrix with positive definite real part. Furthermore, we used the freedom of choice in the phase of the spinors $\xi_{ab}$ to add an additional phase $e^{-i j_{ab}\theta_{ab}}$, whose purpose is to cancel the non-Regge-like phase\footnote{This non-Regge-like phase  depends only on the boundary data and  plays the same role as the phase $e^{-i j_{ab} \Pi_{ab}}$ appearing in the Lorentzian EPRL model \cite{Barrett:2009mw}, where $ \Pi_{ab}$ is zero if both normals of the tetrahedra $\tau_a$ and $\tau_b$ are future or both past pointing and otherwise equals $\pi$. In fact, it is possible to impose a similar restriction (Regge-phase convention) on the phases of $\xi_{ab}$  as in \cite{Barrett:2009mw}, so that  $j_{ab}\theta_{ab}$ reduces to $j_{ab} \Pi_{ab}$.} in the asymptotic limit of the model (see section 14 of \cite{Haggard:2014xoa}). 
With this choice of a boundary state the semiclassical 2-point function is given by 
\begin{equation}
\label{eq:2ptsc}
\begin{split}
G^{abcd}_{mn}=&\frac{\sum_{\vec{j}} \int\mathfrak{D}A\,\mathfrak{D}\bar A\, d\mu(z) \,e^{S_{tot}}\, q^n_{ab}\, q^m_{cd}}{\sum_{\vec{j}} \int\mathfrak{D}A\,\mathfrak{D}\bar A\, d\mu(z) \,e^{S_{tot}}}
-\frac{\sum_{\vec{j}} \int\mathfrak{D}A\,\mathfrak{D}\bar A\, d\mu(z) \,e^{S_{tot}}\, q^n_{ab}}{\sum_{\vec{j}} \int\mathfrak{D}A\,\mathfrak{D}\bar A\, d\mu(z) \,e^{S_{tot}}}\,
\frac{\sum_{\vec{j}} \int\mathfrak{D}A\,\mathfrak{D}\bar A\, d\mu(z) \,e^{S_{tot}}\, q^m_{cd}}{\sum_{\vec{j}} \int\mathfrak{D}A\,\mathfrak{D}\bar A\, d\mu(z) \,e^{S_{tot}}}~,
\end{split}
\end{equation}
where we introduced the short notation:
\beq 
\label{eq:q}
q^n_{ab}:= \delta_{ij} A^i_{na} A^j_{nb}
\eeq
and 
\begin{equation}
\label{eq:actt}
S_{tot}=\left[-\frac{1}{2}\sum_{(ab)(cd)}\gamma\alpha_{(ab)(cd)}\frac{ j_{ab}-j_{0ab}}{\sqrt{ j_{0ab}}}\frac{ j_{cd}- j_{0cd}}{\sqrt{ j_{0cd}}}-i\sum_{(ab)}(\gamma\phi^0_{ab}( j_{ab}- j_{0ab}) - j_{ab}\theta_{ab})\right]+S[A,\bar{A},\vec{z};\vec{j},\vec{\xi}]~.
\end{equation}

\section{Asymptotic Limit of the 2-point function}
\label{sec:lim}

In the following we derive the first non-vanishing contribution of the asymptotic expansion of $G^{abcd}_{mn}(\lambda)$ for large $\lambda$ where $j\mapsto\lambda j$, $h\mapsto \lambda h$ and $\bar h \mapsto \lambda \bar h$. Under this rescaling of the parameters, the action \eqref{eq:actt} scales as $S_{tot}(\lambda)=\lambda S_{tot}(1)$ and $q^n_{ab}$ scales as $\lambda^2 q^n_{ab}$ in the old as well as in the new proposal. If the spins $j$ become large one can furthermore approximate the sums over the spins in \eqref{eq:actt} by integrals using the Euler-Maclaurin formula, so  that 
\begin{equation}
	\label{eq:2ptsct2}
\frac{1}{\lambda^4}G^{abcd}_{mn}(\lambda)\sim
\frac{\int_{\vec{j}} \int\mathfrak{D}A\,\mathfrak{D}\bar A\, d\mu(z) \,e^{\lambda S_{tot}}\, q^n_{ab}\, q^m_{cd}}{\int_{\vec{j}} \int\mathfrak{D}A\,\mathfrak{D}\bar A\, d\mu(z) \,e^{\lambda S_{tot}}}
-\frac{\int_{\vec{j}} \int\mathfrak{D}A\,\mathfrak{D}\bar A\, d\mu(z) \,e^{\lambda S_{tot}}\, q^n_{ab}}{\int_{\vec{j}} \int\mathfrak{D}A\,\mathfrak{D}\bar A\, d\mu(z) \,e^{\lambda S_{tot}}}\,
\frac{\int_{\vec{j}} \int\mathfrak{D}A\,\mathfrak{D}\bar A\, d\mu(z) \,e^{\lambda S_{tot}}\, q^m_{cd}}{\int_{\vec{j}} \int\mathfrak{D}A\,\mathfrak{D}\bar A\, d\mu(z) \,e^{\lambda S_{tot}}}~.
\end{equation}
However, the asymptotics of \eqref{eq:2ptsct2} is only well-defined if the critical points of the integrants are isolated. Therefore, we need to fix the only remaining gauge freedom\footnote{In the action \eqref{eq:act1} this invariance is broken at the vertices $\{v_i\}$ of $\Gamma_5$ due to the gauge fixing \eqref{eq:abc}. But $\{v_i\}$ is a set of measure zero in $\int_{S^3}$ so that a gauge fixing of the Chern-Simons connection can be imposed equivalently on the punctured sphere $S^3\setminus\{v_i\}$.}
under the local transformation $A \to A^{g(x)}= g(x) A g(x)^{-1} + g dg(x)$ by e.g. imposing the gauge fixing condition discussed in \cite{bar-natan1991} on $W[A]$ and $W[\bar{A}]$ (see appendix \ref{app:gauge}). This modifies  the path-integral measure by adding a Faddeev-Popov determinant. But due to the normalization in \eqref{eq:2ptsct2} the dependence on the measure factor drops out, so that the Faddeev-Popov determinant can be safely ignored. Hence, the details of the gauge fixing are not of importance for the following and will be suppressed for the sake of simplicity.


\subsection{Calculation scheme}\label{sbs:cal}
Before the first non-vanishing order  \eqref{eq:2ptsct2} is calculated explicitly, let us briefly explain the reasoning behind the calculation.  
Suppose $S(x,y),x\in\mathbb{R}^{dx},y\in\mathbb{R}^{dy}$ is a complex valued function, and it is smooth in the neighborhood of the point $(\mathring{x},\mathring{y})$,  which is a solution of the critical conditions $Re(S)=0$ and $\delta S=0$. 
Then the asymptotic expansion of the integral ${\cal I}=\int dxdy\ u(x,y)e^{\lambda S(x,y)}$, where $u$ is a function with compact support in a neighborhood of the critical point, can be derived by expanding every integral separately.
More specifically, in the first step, we only expand the integral in $x$ and leave $y$ as free parameters. As the action $S$ is a complex valued function, the number of critical conditions $\delta_{x}S=0$ is twice as many as the number of the type-$x$ variables. So for  $y$ close to $\mathring{y}$, there is no guarantee that $\delta_{x}S=0$ has a solution. However, the almost-analytical extension\footnote{An almost analytic extension of a function is in general not uniquely defined. However, the asymptotic expansion of ${\cal I}$ does not depend on the details of the chosen extension (see e.g \cite{Hoermander, Melin}  for a proof).} of the action has a solution for critical condition $\delta_{\tilde{x}}S=0$, which is denoted as $\tilde{x}(y)$. According to Theorem 2.3 in \cite{Melin}\footnote{ Also see Theorem 7.7.12 in \cite{Hoermander} and \cite{Han:2013hna}} the asymptotic expansion in $x$ is given by
\begin{equation}
\begin{split}\label{eq:ix}
{\cal I} = \int dxdy\  u(x,y) e^{\lambda S(x,y)}
=(\frac{2\pi}{\lambda})^{\frac{dx}{2}}\int dy\,\tilde{u}(y)\, e^{\lambda S(\tilde{x}(y),y)}
\end{split}
\end{equation}
with
\begin{equation}
\label{eq:up}
\begin{split}
\tilde{u}(y)&=\frac{e^{i Ind H_{xx}}}{\sqrt{|det H_{xx}|}}\left[u(\tilde{x},y)+\frac{1}{\lambda}\left(\frac{1}{2}u^{\prime\prime}_{x_ix_j}(\tilde{x},y)H^{-1}_{x_ix_j}+D\right)+\mathcal{O}(\frac{1}{\lambda^2})\right]\Bigg\rvert_{\tilde{x}=\tilde{x}(y)}~.
\end{split}
\end{equation}
Here $H_{xx}$ is Hessian with respect to $x$, $d_x$ is the rank and Ind $H$ is the index of the Hessian $H$, $u^{\prime\prime}_{x_ix_j}=\partial^2 u/\partial x_i\partial x_j$ and $D$ depend linearly on $u$ and its first order derivative $u'_x$ at $\tilde{x}=\tilde{x}(y)$.\footnote{
	Explicitly D is given by 
	\[D(y,z)=u^{\prime}_iR^{\prime\prime\prime}_{jkl}(H_{xx}^{-1})^{ij}(H_{xx}^{-1})^{kl}+\frac{5}{2}uR^{\prime\prime\prime}_{ijk}R^{\prime\prime\prime}_{mnl}(H_{xx}^{-1})^{im}(H_{xx}^{-1})^{jn}(H_{xx}^{-1})^{kl}\rvert_{\tilde{x}=\tilde{x}(y,z)},\]
	where $R(k,y)=S(k,y)-S(\tilde{x},y)-\frac{1}{2}(H_{xx})_{ij}(k-\tilde{x})^i(k-\tilde{x})^j$. See e.g. \cite{Hoermander} for details.}
In the next step the integral of $y$ is expanded around its critical point. 
The almost-analytical extension also grants the existence of a solution of $\delta_{\tilde{y}}S(\tilde{x}(\tilde{y}),\tilde{y})=0$. 

As the almost-analytical extended parameter $\tilde{y}$ can be valued arbitrarily in the neighborhood of $\mathring{y}$, there may be more than one critical point for the system. However, according to theorem 2.3 in \cite{Melin} (or theorem 3.1 in \cite{Han:2013hna}), the only critical point $(\tilde{x}(y),y)$ that also solves $Re(S)=0$ is the critical point on the real axis, i.e. $(\mathring{x},\mathring{y})$. Thus, imposing the reality condition $(\tilde{x}(y),y)=Re(\tilde{x}(y),y)$ in \eqref{eq:hc11} singles out the correct critical point and the asymptotic expansion of \eqref{eq:ix} is given by  
\begin{equation}
	\begin{split}\label{eq:hc11}
	{\cal I}&=\left(\frac{2\pi}{\lambda}\right)^{\frac{dx+dy}{2}}\frac{e^{i (Ind \tilde{H}_{yy}+Ind H_{xx})}}{|det\tilde{H}_{yy}|\cdot|det H_{xx}|}\,e^{\lambda S}\\
	&\cdot\left(u+\frac{1}{2 \lambda}(u^{\prime\prime}_{x_ix_j}H^{-1}_{x_ix_j}+ \tilde{u}^{\prime\prime}_{y_iy_j}\tilde{H}^{-1}_{y_iy_j}+\tilde{D})+\mathcal{O}(\frac{1}{\lambda^2})\right)\Bigg\rvert_{(\tilde{x}(\tilde{y}),\tilde{y})}~.
	\end{split}
\end{equation}
Here, $\tilde{H}$ is the Hessian of the action $\tilde{S}(\tilde{x}(\tilde{y}),\tilde{y})$, $\tilde{u}_{y_i,y_j}$ stands for the second derivative of the function $u(\tilde{x}(\tilde{y}),\tilde{y})$ and $\tilde{D}$ is a function that only depends on $u$ and its first order derivatives.
Note that all the multiplicative prefactors in \eqref{eq:hc11} cancel in the graviton propagator \eqref{eq:2ptsct2}. Moreover a short calculation shows that also all terms that are not proportional to $(q^n_{ab})' (q^m_{cd})'$ drop out so that finally
\begin{widetext}
	\begin{equation}
	\label{eq:hd}
	\begin{split}
	\frac{1}{\lambda^4}G(\lambda)=&\frac{\int\ dxdy\, q(x,y)\,p(x,y)\,e^{\lambda S(x,y)}}{\int\ dxdy \,e^{\lambda S(x,y)}}-\frac{\int\ dxdy\, q(x,y)\,e^{\lambda S(x,y)}}{\int\ dxdy \,e^{\lambda S(x,y)}}\frac{\int\ dxdy\, p(x,y)\,e^{\lambda S(x,y)}}{\int\ dxdy \,e^{\lambda S(x,y)}}\\
	=&\frac{1}{\lambda}\left[p^\prime_{x_i}q^\prime_{x_j}H^{-1}_{x_ix_j}+\tilde{p}^\prime_{y_i}\tilde{q}^\prime_{y_j}\tilde{H}^{-1}_{y_iy_j}\right]+\mathcal{O}(\frac{1}{\lambda^2}).
	\end{split}
	\end{equation} 
\end{widetext}
Thus, we only need to determine the critical point and calculate the inverse Hessians $H^{-1}$ and $\tilde{H}^{-1}$ in order to obtain the asymptotics of $G(\lambda)$. The $x$-variables  represent $z,\bar{z}$ and the fields $A, \bar{A}$ and the role of the $y$-variables is taken by the spins $j$.

\subsection{First non-vanishing order of the graviton propagator}\label{sbs:Propagator}

As mentioned in the previous section, we only need to compute the critical point of the action, the Hessians and the derivatives of \eqref{eq:q}. The critical points and Hessians are obviously independent from the concrete implementation of the metric, only the derivatives of  the  $q$'s depend on this choice. The fix points with respect to connection  and $z$-variables has been calculated in \cite{Haggard:2014xoa} (see appendix \ref{app:cpoint}). As proven in \cite{Haggard:2014xoa}, the path integral is peaked on flat connections with defects associated to the edges of the graph $\Gamma^5$ (see Fig. \ref{fig:g5}). These can be interpreted as describing the parallel transport along the edges/faces of a constantly curved 4-simplex, where the sign of the curvature is determined by the boundary data. Furthermore, the action $S[A,\bar{A},\vec{z};\vec{j},\vec{\xi}]$ in  \eqref{eq:actt} reduces to the Regge-action at the critical point. So it only remains to calculate the critical point with respect to the spins. It is easy to see that the real part of \eqref{eq:actt} only vanishes if $j_{ab}=j^0_{ab}$. Recall that in the general scheme, discussed above, the critical conditions with respect to $j$ are enforced after the critical conditions for $z$ and $A$ have been evaluated. Thus the action $S[A,\bar{A},\vec{z};\vec{j},\vec{\xi}]$ in  \eqref{eq:actt} reduces to the Regge-action and we obtain 
\begin{equation}
\label{eq:variationj}
	\frac{\partial S_{total}}{\partial j_{ab}}\bigg\rvert _{crit}=-i\gamma\phi^{ab}_0+\frac{\partial S_{Regge}}{\partial j_{ab}}=0.
\end{equation}
The Hessians are straight forward to calculate and the explicit expressions are given in appendix \ref{app:Hessian}. In the standard proposal, where the metric operator is defined through the canonical one, $A^i_{ab}$ equals $\gamma j_{ab}\frac{\scal{\sigma^iz_{ab},\xi_{ba}}}{\scal{z_{ab},\xi_{ba}}}$ (see section \ref{sec:2-p}) which implies
\begin{equation}
\begin{split}\label{eq:del}
&\delta_{\bar{z}_{na}}q^n_{ab}=\gamma^2j_{0na}j_{0nb}\left[\frac{e^{i\phi_{an}\sigma^i\xi_{an}}}{|z_{na}|}-\frac{e^{i\phi_{an}}\xi_{an}}{|z_{na}|}n^i_{an}\right]n^i_{bn}\\
&\delta_{z_{an}}q^{n}_{ab}=0\\
&\delta_{A(x)}q^n_{ab}=0\\
&\delta_{\bar{A}(x)}q^n_{ab}=0~.
\end{split}
\end{equation}
Thus, only the inverse Hessians $H^{-1}_{\bar{z}_{ad}\bar{z}_{cd}}$ and $H^{-1}_{j_{ad}j_{cd}}$ are needed. Even though $H_{\bar{z}_{ad}\bar{z}_{cd}}=0$ the inverse does not vanish since the Hessian is not block diagonal in $z$ and $A$-variables (see equations \eqref{eq:zz}-\eqref{eq:BABA}). However, $H_{z\bar{z}},\, H_{zA}, \, H_{z\bar{A}}, \, H_{AA}$ etc. are all of the form $\lambda(\hat{H}+\mathcal{O}(\gamma))$, where $\hat{H}$ is independent of $\gamma$ and $\lambda$, while $H_{zz}$ and $H_{\bar{z}\bar{z}}$ vanish. This implies that $H^{-1}_{\bar{z}\bar{z}}$  must be  of the form $\lambda^{-1}(\hat{H}_{\bar{z}\bar{z}}^{-1}+\mathcal{O}(\gamma))$ as well so that 
\bq
\delta_{\bar{z}}q \,\delta_{\bar{z}}q \,H^{-1}_{\bar z \bar z}=\gamma^4\lambda^3(\hat{H}^{-1}_{\bar z \bar z}+\mathcal{O}(\gamma))~.
\eq 
On the other hand, the type-$y$ Hessian $\tilde{H}_{j_{ab}j_{cd}}$ is given by
\begin{equation}
\label{eq:HJJ}
\delta_{j_{ab}}\delta_{j_{cd}}\bigg\rvert_{crit}S_{tot}=-\frac{\gamma\alpha^{(ab)(cd)}}{\sqrt{j_{0ab}}\sqrt{j_{0cd}}}+\delta_{j_{ab}}\delta_{j_{cd}}\bigg\rvert_{crit}S_{Regge},
\end{equation}
where $S_{Regge}=\sum_{(ab)}\gamma (j_{ab} \Theta_{ab}-\gamma\Lambda V_4)$ and $V_4$ is the 4-volume of the simplex for $\gamma=1$. Consequently, the first non-vanishing order of the graviton propagator with the standard definition of the metric operator is given by 
\begin{equation}
\label{eq:R1}
\begin{split}
G(\lambda)\approx (\tilde{q}^n_{ab})^\prime_{j_{cd}} \,(\tilde{q}^m_{cd})^\prime_{j_{ef}}\,\tilde{H}^{-1}_{j_{cd}j_{ef}} +\gamma^4\lambda^3(\hat{H}+\mathcal{O}(\gamma)).
\end{split}
\end{equation}
The first term  in \eqref{eq:R1} is the expected semiclassical expression and  scales as $\gamma^3\lambda^3$ except for the term proportional to the volume, which scales as $\gamma^4\lambda^3$. The additional second term in \eqref{eq:R1} would be suppressed if we send $\gamma$ to zero in addition to the above limit\footnote{This is the so-called double scaling limit considered in previous calculations \cite{Alesci:2007tx, Alesci:2007tg,Bianchi:2009ri, Bianchi:2011hp, Shirazi:2015hwp} without cosmological constant.}. Unfortunately, this would also suppress the volume term and can therefore not be used to recover the expected result. 

In the alternative metric proposal, discussed in section \ref{sec:propagator}, $q^n_{ab}$ is independent of $\bar{z}$ so that the second term in \eqref{eq:R1} drops out automatically, since in this case $\delta_{\bar{z}}q $ in \eqref{eq:del} vanishes, and the additional limit $\gamma\to0$ can be avoided. This also applies to the standard EPRL model without cosmological constant. To summarize, we observe that we can only recover the expected semiclassical expression of the graviton propagator in the model \cite{Haggard:2014xoa,Haggard:2015ima, Haggard:2015kew,Haggard:2015yda} if the metric operator is defined purely by the areas, which strengthens our view point that a different construction of the graviton propagator might be necessary.

\section{Summary}
\label{sec:dis}

In this paper we explored the semiclassical limit of the metric 2-point function for the model \cite{Haggard:2014xoa,Haggard:2015ima, Haggard:2015kew,Haggard:2015yda}, which includes a cosmological constant. If the metric operator is defined through the flux operators of canonical LQG alike previous works on the graviton propagator, then the first non-vanishing order in the limit $j,|h|\to \infty$ contains an additional term next to the expected semiclassical contribution. In contrast to models without a cosmological constant, this additional term can not be suppressed by taking the limit $\gamma\to 0$ since this limit would also suppress the cosmological constant in the semiclassical action. We therefore suggested to reconsider the definition of the graviton propagator itself. 

Since spin foam models are build from discretized model, it is tempting to interpret them as truncated theories whose true nature becomes transparent only in a continuum limit. Moreover, the 2-point function is a tool deeply rooted in perturbative quantum field theory. In this light, we may interpret the semiclassical boundary states as ``vacuum states'' of a truncated theory around which we are considering quantum perturbations. So each boundary state gives rise to a different truncated theory. One might therefore wonder whether it wouldn't be more appropriate to define the metric operator in a way that is directly adapted to the boundary state and the so defined truncated theory. Since we are working in a discretized setting, the first guess would be to define the metric by the edge lengths of the discretization over which the semiclassical states are peaked. However, this would not give rise to a proper operator as spin foams depend only on the area, i.e. the spins. Even though the edge lengths are not uniquely solvable in terms of the areas, it is still possible to define the metric in terms of the areas since the ambiguity is discrete and can be uniquely fixed through the boundary states. The simplest choice would be to define the tetrad operator as $j n^i_{ab}$, where $n_{ab}$ is the outward pointing normal of the triangle shared by tetrahedron $\tau_a$ and $\tau_b$. This choice might be too simple in the sense that it heavily depends on the boundary data and, hence, might suppress interesting quantum fluctuations. Yet, area and normals to the faces heavily overdetermine the 4-simplex so a more quantum definition of the metric operator might be found, whose dependence on the boundary state is less restrictive. In any case if the metric operator only depends on the spins and no other variables of the path integral then the second nonclassical term in the propagator is no longer present and the expected semiclassical result is recovered. This would also supersede the limit $\gamma\to 0$ in models without cosmological constant.   

To conclude, it seems to be necessary to redefine the graviton propagator by considering a more truncated scenario in order to recover the semiclassical expression in models with cosmological constant. Of course, the considerations here are only valid for a single 4-simplex and a final conclusion should be only drawn after implementing a continuum limit. This is a research undergoing and will be reported elsewhere.

\section*{Acknowledgements}
AZ acknowledges support by a Feodor Lynen Research Fellowship of the Alexander von Humboldt-Foundation. Zichang Huang appreciate the useful discussion with Hongguang Liu.  MH acknowledges support from the US National Science Foundation through grant PHY-1602867, and startup grant at Florida Atlantic University, USA.

\appendix
\section{The area of the triangle in the geometry with a cosmological constant}\label{app:area}

As it is pointed out in \cite{Haggard:2015ima}, all faces of the curved 4-simplex are flatly embedded in the ambient space $\mathcal{S}_3$ or $\mathcal{H}_3$. 

In the spherical case this means each of the triangle faces of the 4-simplex is a part of the great 2-sphere of $\mathcal{S}_3$ and it is enclosed by the edges which are the great cycles of the 2-sphere. Geometrically this kind of triangle is called a spherical triangle.  The area of a spherical triangle is given by the equation
\begin{equation}
D=R^2(A+B+C-\pi)
\end{equation}
where $A,B,C$ are the interior angles of the triangle. The radius of the great 2-sphere can be given by the cosmological constant through $R=1/\sqrt{\Lambda}$. By redefining the unit properly we can normalize the radius to $1$. By using the spherical sine law and spherical cosine law, one can get an expression of the area in terms of the edge-lengths. In chapter.\uppercase\expandafter{\romannumeral8} of the book \cite{booksptri}, this is given by the L'Huilier's theorem
\be\label{eq:sparea}
\tan{\frac{D}{4R^2}}=\sqrt{\tan{\half s}\tan{\half(s-a)}\tan{\half(s-b)}\tan{\half(s-c)}},
\ee
where $a,b,c$ are edge lengths and $s=\half(a+b+c)$.

For the hyperbolic case, the area is given as
\be
D=R^2(\pi-A-B-C)
\ee
where $A,B,C$ are still interior angles, $R^2$ is given by $R^2=1/\Lambda$. Similarly after choosing a proper unit to normalize the cosmological constant, the hyperbolic sine law and hyperbolic cosine law can convert the area as
\be\label{eq:hyparea}
\tan(\frac{D}{2R^2})=\frac{\sqrt{1-\cosh^2{a}-\cosh^2{b}-\cosh^2{c}+2\cosh{a}\cosh{b}\cosh{c}}}{1+\cosh{a}+\cosh{b}+\cosh{c}},
\ee 
where $a,b,c$ are edge lengths \cite{2011arXiv1102.0462B}.

For a 4-simplex the sign of the determinant of the tetrahedron gram matrix can sufficiently show whether it is spherical geometry, hyperbolic geometry or flat geometry \cite{Haggard:2015ima}. The elements of the gram matrix are the cosine function of the 2d dihedral angles which are given by the boundary data. So from a fixed boundary data, one can uniquely decide which equation among \eqref{eq:heron}, \eqref{eq:sparea} and \eqref{eq:hyparea} is needed to construct the metric in terms of the areas as the scheme we discussed in section \ref{sec:propagator}. Small perturbation of the boundary data doesn't change the sign of $\L$ in each tetrahedron. So the metric operator is essentially defined for a neighborhood of boundary data, consistent with the proposal in section \ref{sec:propagator}. 

\section{Calculation of the saddle point and the Hessian}
\label{app:cpoint}

On the critical point the conditions $Re(S)=0$ and $\delta S=0$ are satisfied. For the action  \eqref{eq:actt}, $Re(S_{tot})=0$ is equivalent to
\begin{equation}
\label{eq:crite1}
\begin{split}
-\frac{1}{2}\sum_{(ab)(cd)}\gamma\alpha_{(ab)(cd)}\frac{ j_{ab}-j_{0ab}}{\sqrt{ j_{0ab}}}\frac{ j_{cd}- j_{0cd}}{\sqrt{ j_{0cd}}}+\sum_{(ab)} 2j_{ab} \ln{\frac{<J\xi_{ab},\GI{ab}z_{ab}><z_{ab},\xi_{ba}>}{<\GI{ab}z_{ab},\GI{ab}z_{ab}>^{\frac{1}{2}}<z_{ab},z_{ab}>^{\frac{1}{2}}}}=0
\end{split}
\end{equation}
The first term in \eqref{eq:crite1} is  quadratic which has its maximum value $0$ at $j=j_0$.
The second term in \eqref{eq:crite1} is less than or equal to $0$ due to the Cauchy-Schwartz inequality. Thus, all terms have to vanish separately, which forces $j_{0ab}=j_{ab}$,  $J\xi_{ab}\propto z_{ab}$ and $\xi_{ba}\propto z_{ba}$. Since $\xi$ is normalized this implies
\begin{equation}
\label{eq:critsol}
\begin{split}
J\xi_{ab}=\frac{e^{i\phi_{ab}}}{|\GI{ab}z_{ab}|}& \GI{ab}z_{ab}\,\text{ and }\,\xi_{ba}=\frac{e^{i\phi_{ba}}}{|z_{ab}|} z_{ab}~.
\end{split}
\end{equation}
Combining the two equations yields
\begin{equation}
\label{eq:critsol1}
\xi_{ab}=-e^{i(\phi_{ab}-\phi_{ba})}\frac{|z_{ab}|}{|\GI{ab}z_{ab}|}G_{ab}J\xi_{ba}
\end{equation} 
where we used $JG_{ab}^{-1}=G_{ab}^\dagger J$ and $J^{-1}=-J$.
The condition $\delta S_{total}=0$ stands for the derivatives w.r.t the four types of variables, $z,A$ and $j$ respectively. 
The derivative w.r.t $z$ is\footnote{Since $z$ is an element on a Riemann surface $\mathbb{CP}_1$, the variation of $z$ is perpendicular to $z$ itself, i.e. $\delta z= \epsilon Jz$.}, 
\begin{widetext}
	\begin{equation}
	\label{eq:varsz}
	\begin{split}
	\delta_{(z_{ab})} S_{total}
	=&\epsilon\left[2j_{ab}\frac{<J\xi_{ab},\GI{ab}Jz_{ab}>}{<J\xi_{ab},\GI{ab}z_{ab}>}-j_{ab}(1+i\gamma)\frac{<(G_{ab}^\dagger G_{ab})^{-1} z_{ab},Jz_{ab}>}{<(G_{ab}^\dagger G_{ab})^{-1}z_{ab},z_{ab}>}\right]\\
	+&\bar\epsilon\left[2j_{ab}\frac{<Jz_{ab},\xi_{ba}>}{<z_{ab},\xi_{ba}>}-j_{ab}(1+i\gamma)\frac{<Jz_{ab}, (G_{ab}^\dagger G_{ab})^{-1} z_{ab}>}{<(G_{ab}^\dagger G_{ab})^{-1}z_{ab},z_{ab}>}\right],
	\end{split}
	\end{equation}
\end{widetext}
which splits into two independent equation for $\epsilon$ and $\bar\epsilon$.
Equation \eqref{eq:critsol} implies $<Jz_{ab},\xi_{ba}>=0$, which means that $<Jz_{ab}, (G_{ab}^\dagger G_{ab})^{-1} z_{ab}>=<(G_{ab}^\dagger G_{ab})^{-1} z_{ab},Jz_{ab}>$ has to vanish as well.
 But this implies in turn that
\begin{equation}\label{eq:varz2}
<J\xi_{ab},\GI{ab}Jz_{ab}>=0,
\end{equation} 
in the term proportional to $\epsilon$.
Again by using that the $\xi_{ab}$'s are normalized and by using \eqref{eq:critsol}, we find that
\begin{equation}
\label{eq:varz1}
\begin{split}
&\frac{e^{-i\phi_{ab}}}{|\GI{ab}z_{ab}|}<\GI{ab}z_{ab},J\xi_{ab}>=\frac{e^{-i\phi_{ba}}}{|z_{ab}|}<z_{ab},\xi_{ba}>\\
\Leftrightarrow&<z_{ab},G^{-1}_{ab}J\xi_{ab}>=\frac{|\GI{ab}z_{ab}|}{|z_{ab}|}e^{i(\phi_{ab}-\phi_{ba})}<z_{ab},\xi_{ba}>.
\end{split}
\end{equation}
Because $z_{ab}$ and $Jz_{ab}$ are orthogonal, \eqref{eq:varz1} and \eqref{eq:varz2} combine into
\begin{equation}
\label{eq:critsol3}
\begin{split}
&J\xi_{ab}=e^{i(\phi_{ab}-\phi_{ba})}\frac{|\GI{ab}z_{ab}|}{|z_{ab}|}G_{ab}\xi_{ba}. 
\end{split}
\end{equation}
The derivatives with respect to $A$ and $\bar{A}$ can be calculated by using the technique of \cite{Haggard:2014xoa}. We find:
\begin{widetext}
	\begin{equation}
	\label{eq:dera}
	\frac{\delta S_{tot}}{\delta A^i_{\mu}(x)}|_{Re(S_{tot}=0)}= \frac{ih}{16\pi}\epsilon^{\mu\nu\rho}F^i_{\nu\rho}[A(x)]-(1+i\gamma)j_{cd}<\left[(G_{c,s_0}^{-1})^\dagger\tau_iG_{c,s_0}^\dagger\right]J\xi_{cd},J\xi_{cd}>\delta^{(2)\mu}_{\ \ \ l_{cd}}(x)=0\ \forall x\in l_{cd} 
	\end{equation}
\end{widetext}
and
\begin{widetext}
	\begin{equation}
	\label{eq:derb}
	\frac{\delta S_{tot}}{\delta \bar{A}^i_{\mu}(x)}|_{Re(S_{tot}=0)}= \frac{i\bar{h}}{16\pi}\epsilon^{\mu\nu\rho}\bar F^i_{\nu\rho}[\bar A(x)]+(1-i\gamma)j_{cd}<J\xi_{cd},\left[\GI{c,s_0}\tau_iG_{c,s_0}^\dagger\right]J\xi_{cd}>\delta^{(2)\mu}_{\ \ \ l_{cd}}(x)=0\ \forall x\in l_{cd}~,
	\end{equation}
\end{widetext}
where the $\delta$-function is defined by
\begin{equation}
\label{eq:delta2}
\delta^{(2)\mu}_{\ \ \ l}(x)=\int_{0}^{1}\delta^{(3)}(x-l(s))\frac{dl^\mu}{d s} ds.
\end{equation}
Equations \eqref{eq:dera} and \eqref{eq:derb} imply that the $SL(2,\mathbb{C})$-connection is flat on $S_3$ except on the edges of the $\Gamma_5$ graph (see Fig. \ref{fig:g5}). 
This can be related to the closure condition for curved tetrahedra \cite{Haggard:2014xoa}.
The last variation with respect to $j$ yields
\begin{widetext}
	\begin{equation}
	\label{eq:variationj2}
	\begin{split}
	\frac{\partial S_{total}}{\partial j_{ab}}\bigg\rvert _{crit}=&\left[-\sum_{(cd)}\frac{\gamma\alpha^{(ab)(cd)}(j_{cd}-j_{0cd})}{\sqrt{j_{0cd}}\sqrt{j_{0ab}}}-i(\gamma\phi^{ab}_0-\theta^{ab})+\frac{\partial S}{\partial j_{ab}}\right]\Bigg\rvert_{crit}=-i\gamma\phi^{ab}_0+\frac{\partial S_{Regge}}{\partial j_{ab}}=0.
	\end{split}
	\end{equation}
\end{widetext}
In the last step we used that the action  \eqref{eq:act1} evaluated at the critical points reduces to the Regge-action with a cosmological constant plus a non-Regge like phase (see \cite{Haggard:2014xoa} for details). Since the non-Regge like phase only depends on the boundary data the phase $\theta^{ab}$ can be chosen in such a way that it cancels the non-Regge like phase leading to the above result.

\section{Hessians}
\label{app:Hessian}
The Hessians at the critical point are given by:
\begin{align}
\label{eq:zz}
H_{\bar{z}_{cd}\bar{z}_{ef}}
=&0,
\\
\label{eq:bzbz}
H_{z_{ef}z_{cd}}
=&0,
\\
\label{eq:zbz}
H_{z_{cd}\bar{z}_{ef}}=&\delta^{(ef)(cd)}j_{ef}\left[\frac{(i\gamma-1)}{|z_{ef}|^2}-\frac{(i\gamma+1)G_{ef}^{-1}\GI{ef}}{|\GI{ef}z_{ef}|^2}\right],
\\
\label{eq:za}
H_{z_{ef}A^i_{\mu}(x)}&=\frac{-(1+i\gamma)j_{ef}e^{i\phi_{ef}}}{|\GI{ef}z_{ef}|}\left[(G_{ef}^{-1}\GI{e s_0}\tau_iG_{es_0}^\dagger J\xi_{ef})^\dagger-<J\xi_{ef},G_{es_0}\tau_iG_{es_0}^{-1} J\xi_{ef}>(G_{ef}^{-1}J\xi{ef})^\dagger\right]\delta^\mu_{l_{ef}}(x),
\\
\label{eq:bza}
H_{\bar{z}_{ef}A^i_{\mu}(x)}&=\frac{-(1+i\gamma)j_{ef}e^{-i\phi_{ef}}}{|\GI{ef}z_{ef}|}\left[(G_{s_0f}^{-1}\tau_iG_{es_0}^{-1}J\xi_{ef})-<J\xi_{ef},G_{es_0}\tau_iG_{es_0}^{-1} J\xi_{ef}>(G_{ef}^{-1}J\xi_{ef})\right]\delta^\mu_{l_{ef}}(x),
\\
\label{eq:zba}
H_{z_{ef}\bar{A}^i_{\mu}(x)}&=\frac{(1-i\gamma)j_{ef}e^{i\phi_{ef}}}{|\GI{ef}z_{ef}|}\left[(G_{s_0 f}^{-1}\tau_i G_{e s_0}^{-1}J\xi_{ef})^\dagger-<J\xi_{ef},\GI{es_0}\tau_iG_{es_0}^\dagger J\xi_{ef}>(G_{ef}^{-1}J\xi_{ef})^\dagger\right]\delta^\mu_{l_{ef}}(x),
\\
\label{eq:bzba}
H_{\bar{z}_{ef}\bar{A}^i_{\mu}(x)}&=\frac{-(1+i\gamma)j_{ef}e^{-i\phi_{ef}}}{|\GI{ef}z_{ef}|}\left[(G_{ef}^{-1}\GI{e s_0}\tau_iG_{es_0}^{\dagger}J\xi_{ef})-<J\xi_{ef},\GI{es_0}\tau_iG_{es_0}^\dagger J\xi_{ef}>(G_{ef}^{-1}J\xi_{ef})\right]\delta^\mu_{l_{ef}}(x),
\end{align}
where $s_0$ stands for a point on the edge $l_{ef}$ with a coordinate $x$.
Furthermore, 
\begin{equation}
\label{eq:baa}
\begin{split}
H_{\bar{A}^j_\nu(y) A^i_\mu(x)}&=(1+i\gamma)j_{cd}\left[<(\GI{es_0}\tau_iG_{es_0}^\dagger)J\xi_{ef}, (\GI{es_0^\prime}\tau_jG_{es_0^\prime}^\dagger) J\xi_{ef}>\right.\\
&\left.-<J\xi_{ef},(\GI{es_0}\tau_iG_{es_0}^\dagger)J\xi_{ef}><J\xi_{ef},(\GI{es_0^\prime}\tau_jG_{es_0^\prime}^\dagger)J\xi_{ef}>\right]\delta^\mu_{l_{ef}}(x)\delta^\nu_{l_{ef}}(y),
\end{split}
\end{equation}
where $s_0^\prime$ stands for a $y$ point on the edge $l_{ef}$.
The Hessian element $H_{A^i_\mu(x_1)A^j_\nu(x_2)}$ and $H_{\bar{A}^i_\mu(x_1)\bar{A}^j_\nu(x_2)}$ are 
\begin{equation}
\label{eq:AA}
\begin{split}
H_{A^j_\nu(y) A^i_\mu(x)}=&\frac{ih}{16\pi}\epsilon^{\mu\lambda\rho}(\delta_{A^j_\nu(y)}F^i_{\lambda\rho}[A(x)])\\
-&(1+i\gamma)j_{ab}\left[<\mathcal{P}(\GI{as_0}\tau_jG\GI{s_0s_0^\prime}\tau_iG_{as_0^\prime}^\dagger) J\xi_{ab},J\xi_{ab}>\right.\\
-&\left.<\GI{as_0}\tau_jG_{as_0}^\dagger J\xi_{ab},J\xi_{ab}><\GI{as_0^\prime}\tau_iG_{as_0^\prime}^\dagger J\xi_{ab},J\xi_{ab}>\right]\delta^\mu_{l_{ab}}(x)\delta^\nu_{l_{ab}}(y)
\end{split}
\end{equation}

\begin{equation}
\label{eq:BABA}
\begin{split}
H_{\bar{A}^j_\nu(y) \bar{A}^i_\mu(x)}&=\frac{i\bar{h}}{16\pi}\epsilon^{\mu\nu\rho}(\delta_{\bar{A}^j_\nu(y)}\bar F^i_{\nu\rho}[\bar A(x)])\\
&+(1-i\gamma)j_{ab}\left[<J\xi_{ab},\mathcal{P}(\GI{as_0}\tau_i\GI{s_0s_0^\prime}\tau_jG_{as_0}^\dagger)J\xi_{ab}>\right.\\
&-\left.<J\xi_{ab},\GI{as_0}\tau_iG_{as_0}^\dagger J\xi_{ab}><J\xi_{ab},(\GI{as_0^\prime}\tau_jG_{as_0^\prime}^\dagger)J\xi_{ab}>\right]\delta^\mu_{l_{ab}}(x)\delta^\nu_{l_{ab}}(y),
\end{split}
\end{equation}
where $\mathcal{P}$ stands for the path order on the edge $l_{ab}$.
    Note that the contributions from the Chern-Simons term, i.e. $\delta_{A^j_\nu(y)}F^i_{\lambda\rho}[A(x)]$ and $\delta_{\bar{A}^j_\nu(y)}\bar F^i_{\nu\rho}[\bar A(x)]$, are degenerated as long as the remaining gauge freedom under the local transformation $A \to A^{g(x)}= g(x) A g(x)^{-1} + g dg(x)$ is not fixed. Therefore,  we have to  impose a gauge fixing e.g. by following \cite{bar-natan1991}.

\section{Chern-Simons Propagator for Non-Compact Gauge Group}
\label{app:gauge}
In order to get a non-degenerate propagator for the $SL(2,\mathbb{C})$ Chern-Simons action, all gauges need to be fixed. Fortunately, the gauge fixing procedure of Chern-Simons theory for a non-compact group was already derived in the1990s by Bar-Natan and Witten \cite{bar-natan1991}. This gauge fixes the maximum compact subgroup of $SL(2,\mathbb{C})$ in order to obtain a positive definite Hermitian inner product. 

Assume the connection can be expressed as $A=A^{0}+B$, where $A^{0}$ is the critical value, and introduce the following gauge fixing term  
\begin{equation}
\label{eq:gaugefix}
V=\frac{h}{4\pi}\int d\mu\, \Tr[\bar{c}\ast D^{(0)} \ast TB],
\end{equation}
where $\mu$ is Riemann measure on the mannifold defining Chern-Simons theory, $\ast$ is the Hodge star and $T$ is a projector that projects $\mathfrak{sl}(2,\mathbb{C})$  onto  $\mathfrak{su}(2)$. Moreover, $\bar c$ is a Lagrange multiplier of the gauge fixing $D^{(0)}TB=0$. With this gauge fixing the Chern-Simons action can then be written as \footnote{Up to the first order}
\begin{widetext}
	\begin{equation}
	\label{eq:gaugefix1}
	\begin{split}
	S_{CS}=&-i\frac{h}{2}W[A]+\frac{h}{4\pi}\int Tr(-\frac{i}{2}B\wedge D^{(0)}B-i\phi\ast D^{(0)}T\ast B+\bar c\ast D^{(0)}T\ast D^{(0)}c)\\
	&-i\frac{\bar h}{2}W[\bar A]+\frac{\bar h}{4\pi}\int Tr(-\frac{i}{2}\bar B\wedge \bar{D}^{(0)}\bar B-i\tilde{\phi}\ast \bar{D}^{(0)}T\ast \bar{B}+\bar{\tilde{c}}\ast \bar{D}^{(0)}T\ast \bar{D}^{(0)}\tilde{c})\\
	=&-i\frac{h}{2}W[A]-\frac{1}{2}\int Tr(-\frac{i}{2}B^\prime\wedge D^{(0)}B^\prime+2i\phi^\prime\ast TD^{(0)}\ast TB^{\prime})+\int Tr(\bar c^\prime D^{(0)}\ast T D^{(0)}c^\prime)\\
	&-i\frac{\bar h}{2}W[\bar A]-\frac{1}{2}\int Tr(-\frac{i}{2}\bar B^\prime\wedge \bar{D}^{(0)}\bar B^\prime+2i\tilde{\phi}^\prime\ast T\bar{D}^{(0)}\ast T\bar{B}^\prime)+\int Tr(\bar{\tilde{c}}^\prime\bar{D}^{(0)} \ast T \bar{D}^{(0)}\tilde{c}^\prime).\\
	\end{split}
	\end{equation}
\end{widetext}
Here, $B^\prime=B\sqrt{h/4\pi}$ is an $\mathfrak{sl}(2,\mathbf{C})$ algebra valued 1-form, $c^\prime=c\sqrt{h/4\pi}$, $\tilde{c}^\prime=\tilde{c}\sqrt{h/4\pi}$, $\bar c^\prime=\ast \bar c\sqrt{h/4\pi}$ and $\bar{\tilde{c}}^\prime=\ast\bar{\tilde{c}}\sqrt{h/4\pi}$ are $\mathfrak{sl}(2,\mathbf{C})$ algebra valued functions, and $\phi^\prime=T\phi\sqrt{h/4\pi}$ and $\tilde{\phi}^\prime=T\tilde{\phi}\sqrt{h/4\pi}$ are $\mathfrak{sl}(2,\mathbf{C})$ algebra valued 3-forms.
Following \cite{bar-natan1991}, we can now introduce a positive definite  scalar product for the P-forms $u$ and $v$ as
\begin{equation}
\label{eq:pdin}
(u,v)=-\int Tr(u\wedge\ast Tv).
\end{equation}
Define $H:=(B^\prime,\phi^\prime)\in\Omega^1\oplus\Omega^3$ and let $\tilde{H}(\bar{B}^\prime,\tilde{\phi}^{\prime})$ denote it's conjugate then 
\begin{equation}
\label{eq:actiongf}
\begin{split}
S_{CS}=&-i\frac{h}{2}W[A]+\frac{i}{2}(H,\hat{L}_-H)-(\bar c^\prime,\hat{\Delta}_0 c^\prime)\\
&-i\frac{\bar h}{2}W[\bar A]+\frac{i}{2}(\tilde H,\hat{L}_-\tilde H)-(\bar{\tilde{c}}^\prime,\hat{\Delta}_0 \tilde{c}^\prime),
\end{split}
\end{equation}
where $\hat{L}_-=(\ast TD^{(0)}+D^{(0)}\ast T)J$, $\hat{\Delta}_0=\ast T D^{(0)}\ast T D^{(0)}$ and where $J$ equals $-1$, if it acts on a function or 3-form, and equals $1$, if it acts  on a 1-form or a 2-form.
Effectively, this promotes $H$ and $\tilde H$ to the new variables of our theory. The first component of $H$ are just $A$ or $\bar A$ for $\tilde H$, which means that for those components $\delta_H S=0$ or $\delta_{\tilde H}S=0$ impose the critical conditions  \eqref{eq:dera} and \eqref{eq:derb}. 
The derivative with respect to $\phi$ or $\tilde \phi$ yield the gauge fixing condition $D^{(0)}\ast T B^\prime=0$ and $D^{(0)}\ast T \bar B^\prime=0$.

As it is mentioned in \eqref{eq:abc}, we have already chosen a gauge on each vertex to absorb $g_a$ into the holonomy $G_{ab}$. However, this does not impose any restrictions since we integrate over $\mathcal{S}_3$ and the set of vertices 
$\Gamma_5$ is of measure zero so that we can as well integrate over 
$\mathcal{S}_3/\text{vertices}$. Therefore, the gauge fixing and the gauge choice on each vertex can be done at the same time and will not conflict with each other.

\end{document}